\documentclass[lettersize,journal]{IEEEtran}
\usepackage{amsmath,amsfonts}
\usepackage{algorithmic}
\usepackage{algorithm}
\usepackage{array}
\usepackage[caption=false,font=normalsize,labelfont=sf,textfont=sf]{subfig}
\usepackage{textcomp}
\usepackage{stfloats}
\usepackage{url}
\usepackage{verbatim}
\usepackage{graphicx}
\usepackage{cite}
\usepackage{booktabs}
\usepackage{xcolor}
\begin{document}

\title{A Deep Automotive Radar Detector using the \textit{RaDelft} Dataset}

\author{{\IEEEauthorblockN{
Ignacio Roldan\IEEEauthorrefmark{1},   
Andras Palffy\IEEEauthorrefmark{3}\IEEEauthorrefmark{1},    
Julian F. P. Kooij\IEEEauthorrefmark{2},
Dariu M. Gavrila\IEEEauthorrefmark{2},
Francesco Fioranelli\IEEEauthorrefmark{1},
Alexander Yarovoy\IEEEauthorrefmark{1}
}                                     
\\
\IEEEauthorblockA{
\IEEEauthorrefmark{1}Microwave Sensing, Signals and Systems (MS3) Group, Department of Microelectronics\\ \IEEEauthorrefmark{2}Intelligent Vehicles (IV) Group, Department of Cognitive Robotics \\ Delft University of Technology, Delft, The Netherlands\\
\IEEEauthorrefmark{3}Perciv AI, 2628 BC, The Netherlands}
  
 \IEEEauthorblockA{ \emph{\{i.roldanmontero,
a.palffy, 
j.f.p.kooij,
d.m.gavrila,
f.fioranelli, 
a.yarovoy\}@tudelft.nl}}
}
\thanks{Received 30 May 2024; revised 11 September 2024; accepted 17 October 2024. Date of publication 23 October 2024; date of current version
1 November 2024. (Corresponding author: \textit{Ignacio Roldan}.)}
\thanks{Ignacio Roldan, Francesco Fioranelli, and Alexander Yarovoy are with the Microwave Sensing, Signals and Systems (MS3) Group, Department of
Microelectronics, Delft University of Technology (TU Delft), 2628 CD Delft,
The Netherlands (e-mail: i.roldanmontero@tudelft.nl; f.fioranelli@tudelft.nl; a.yarovoy@tudelft.nl).}
\thanks{Andras Palffy is with the Microwave Sensing, Signals and Systems (MS3) Group, Department of Microelectronics, Delft University of Technology (TU Delft), 2628 CD Delft, The Netherlands, and also with Perciv AI, 2628 BC Delft, The Netherlands (e-mail: andras.palffy@perciv.ai)}
\thanks{Julian F. P. Kooij and Dariu M. Gavrila are with the Intelligent Vehicles (IV) Group, Department of Cognitive Robotics, Delft University of Technology (TU Delft), 2628 CD Delft, The Netherlands (e-mail: j.f.p.kooij@tudelft.nl; d.m.gavrila@tudelft.nl).}
\thanks{Copyright (c) 2015 IEEE. Personal use of this material is permitted. However, permission to use this material for any other purposes must be obtained from the IEEE by sending a request to pubs-permissions@ieee.org.}
}

\markboth{IEEE TRANSACTIONS ON RADAR SYSTEMS}%
{Shell \MakeLowercase{\textit{et al.}}: A Sample Article Using IEEEtran.cls for IEEE Journals}


\maketitle

\begin{abstract}
The detection of multiple extended targets in complex environments using high-resolution automotive radar is considered. A data-driven approach is proposed where unlabeled synchronized lidar data is used as ground truth to train a neural network with only radar data as input. To this end, the novel, large-scale, real-life, and multi-sensor \textit{RaDelft} dataset has been recorded using a demonstrator vehicle in different locations in the city of Delft. The dataset, as well as the documentation and example code, is publicly available for those researchers in the field of automotive radar or machine perception. The proposed data-driven detector is able to generate lidar-like point clouds using only radar data from a high-resolution system, which preserves the shape and size of extended targets. The results are compared against conventional CFAR detectors as well as variations of the method to emulate the available approaches in the literature, using the probability of detection, the probability of false alarm, and the Chamfer distance as performance metrics. Moreover, an ablation study was carried out to assess the impact of Doppler and temporal information on detection performance. The proposed method outperforms the different baselines in terms of Chamfer distance, achieving a reduction of 77\% against conventional CFAR detectors and 28\% against the modified state-of-the-art deep learning based approaches.
\end{abstract}


\section{Introduction}
\IEEEPARstart{I}{n} the domain of environment sensing technology, radar sensors can provide unique advantages over other sensors. While lidar offers high-resolution imaging capabilities, making it excellent for detailed environmental mapping, radar provides superior performance in adverse weather conditions, such as fog or rain, or in case of low-light conditions \cite{Sezgin2023}. Furthermore, radar can accurately and directly measure objects' velocity via the Doppler effect. All this makes radar a crucial sensor for vehicular autonomy \cite{Bilik2019}.

A notable trend in automotive radar is the shift towards imaging radar, which achieves high angular resolution in both azimuth and elevation by leveraging a larger number of antennas and thus a larger aperture \cite{Sun2020}. 
Furthermore, Neural Networks (NN) and Deep Learning (DL) techniques are increasingly being applied to signal and data processing \cite{Srivastav2023}. These algorithms can excel in multiple steps of the radar signal processing pipeline, such as detection \cite{Brodeski2019, kradar, Cheng2022, Lin2023, SCORP,Gusland2020, Zheng2023}, classification \cite{Palffy2020, Schumann2020, CRUW}, and signal enhancement \cite{Roldan2023}, offering a richer interpretation of radar data. However, their effectiveness relies on extensive and high-quality datasets for training, to accurately identify and react to diverse driving scenarios. 
To the best of our knowledge, there is a lack of suitable public datasets for radar practitioners where ADC-level (Analog-to-Digital Converter) data from large-aperture radars is collected using real vehicles. Therefore, the \textit{first contribution} of this paper is the introduction of \textit{RaDelft}, a large-scale, real-world multi-sensory dataset recorded in various driving scenarios in the city of Delft, which is publicly shared. 

\begin{figure*}[]
\centerline{\includegraphics[width=\linewidth]{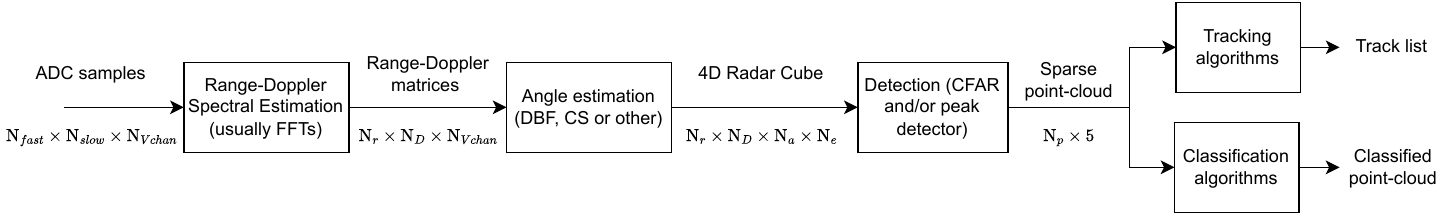}}
\caption{Typical radar processing pipeline, from the raw ADC samples to the output of classification \& tracking steps. $\textrm{N}_{fast}$ and $\textrm{N}_{slow}$ are the number of samples in a chirp and in a CPI, respectively. $\textrm{N}_{Vchan}$ are the number of virtual channels, in a MIMO system the product of the number of Tx and Rx channels. $\textrm{N}_{r}$,$\textrm{N}_{D}$, $\textrm{N}_{a}$, and $\textrm{N}_{e}$ are the number of range, Doppler, azimuth and elevation cells. Finally, $\textrm{N}_{p}$ is the number of points after the detector, with the three spatial coordinates plus Doppler and power.}
\label{RadarPipeline}
\end{figure*}

In terms of signal processing, challenges remain for the integration of radar technology into automotive systems. A primary hurdle in this context is the use of the well-known Constant False Alarm Rate (CFAR) detectors for generating radar point clouds from the dense radar data cube. While CFAR detectors have proven optimal in other environments \cite{Richards2015}, their application in the dynamic and unpredictable conditions of road traffic scenarios suffers from poor performance \cite{Cheng2022,Yoon2019}. Namely, they are designed to maintain a constant rate of false alarms amidst varying clutter, but they struggle to adapt to the rapidly changing environments typical of roadways. Complications such as non-uniform clutter (or the lack of reliable clutter models for this task), target masking, and shadowing can significantly reduce the effectiveness of CFAR detectors in automotive radar settings. Additionally, CFAR detectors are constrained by a fundamental limitation: they typically assume a fixed, expected target size based on predefined guard and training cell hyperparameters. However, in an automotive context, this assumption is problematic as the size of potential targets can widely vary, ranging from medium-sized objects such as pedestrians to large vehicles like trucks or buses. Moreover, the perceived size of these targets in the radar's angular dimension changes with distance. Large objects occupying multiple cells at close range can appear as simpler point-like targets at further distances. This relationship between angular target size and distance adds another layer of complexity to using CFAR detectors in automotive radar, necessitating alternative solutions to accurately detect and classify objects under varying road conditions. 

To address these limitations, the \textit{second contribution} of this work is to present a new data-driven radar target detector using a unique cross-sensor supervision pipeline. The proposed data-driven detector is initially trained with synchronized radar and lidar data together, and can subsequently generate denser point clouds using only raw data from a high-resolution automotive radar. The proposed approach is extensively validated using the aforementioned \textit{RaDelft} dataset.

Compared to the initial results presented in our conference submission \cite{roldan2024cfar}, two additional contributions are presented in this work. First, the proposed data-driven radar detector is expanded to include temporal information across frames, and a more rigorous analysis of the impact of each processing block is included. Second, the multi-sensor dataset used for validation, \textit{RaDelft}, is presented and shared with the broader research community, including example code for easier utilization \cite{RaDelft}.

The rest of the paper is organized as follows. As automotive radar is part of a wider multidisciplinary field on autonomous vehicles, clarifying the terminology used in this work is important to prevent confusion. This is done in Section \ref{terminology}, which also briefly reviews the conventional radar processing pipeline.
Section \ref{relatedWork} reviews the available automotive radar datasets and summarises the state-of-the-art of automotive radar detectors. Section \ref{RaDelftSection} introduces our new publicly available dataset \textit{RaDelft}, detailing its characteristics for data-driven approaches. Our proposed data-driven detector is presented in Section \ref{datadrivendetector}. Section \ref{resultsSection} shows the results of the proposed method and compares them with those of conventional CFAR detectors. Finally, Section \ref{conclusionsSection} concludes the paper.

\section{Terminology \& Radar Processing Review}
In this Section, the terminology used in this work is first clarified, followed by a brief review of the conventional radar processing pipeline and its steps.
\label{terminology}
\subsection{Terminology}
In recent years, automotive radar has become part of a wider multidisciplinary field in autonomous vehicles where scientists from different backgrounds are cooperating. As different research communities might use different terms \cite{Yi2024, Palffy2020}, a list of definitions used in this work is provided here.
\begin{itemize}
  \item \textit{Raw radar data} or \textit{ADC data} refers to the complex baseband samples the ADC provides at each receiver channel. 
  \item \textit{Virtual channel} or \textit{channel} refers to one of the multiple unique combinations of Tx-Rx antenna in a MIMO radar, meaning the signal transmitted from a Tx is received, down-converted, and sampled at the Rx.
  \item \textit{Radar frame} refers to the set of ADC samples from a Coherent Processing Interval (CPI) of each virtual channel. It has dimensions of $\textrm{N}_{fast} \times \textrm{N}_{slow} \times \textrm{N}_{Vchan}$, where these are the number of samples in fast time, number of samples in slow time, and number of virtual channels, respectively.
  \item \textit{Radar cube} refers to the spherical coordinate, discretised representation of the radar data, meaning the range, azimuth, elevation, and Doppler estimation have already been performed. Each cell in the \textit{radar cube} contains a scalar value indicating the reflected power in that cell. The size of each cell is related to the characteristics of the radar, such as the transmitted bandwidth or the antenna array topology. In general, the cells do not have the same size over the whole grid.
  \item An \textit{extended target} is a target occupying multiple cells in one or several dimensions, in contrast to a \textit{point target}, which occupies a single cell. Point targets present a clear peak in the estimation space (range-Doppler-angle), while extended targets do not. 
  \item \textit{Detection} is the binary decision problem determining whether a \textit{radar cube} cell contains only noise or noise plus target. On the other hand, \textit{classification} aims to associate a class to each detected cell, such as 'pedestrian', 'vehicle', or 'light pole', and so on. In general, these two tasks are treated as two blocks in a conventional radar processing pipeline.
  \item \textit{3D occupancy grid} refers to a binary cube, also in spherical coordinates, which contains ones in voxels that are occupied by detected targets, and zeros otherwise. Such a \textit{3D occupancy grid} could be generated directly from a lidar point cloud, but also from a \textit{radar cube} through a detector as this work aims to. In the latter case, the resulting \textit{3D occupancy grid} and the \textit{radar cube} share the same grid.
  \item \textit{Point cloud} refers to a set of $\textrm{N}_p$ points, each containing $\textrm{L}$ features that result from selecting only those cells containing ones in a \textit{3D occupancy grid} and converting them to Cartesian coordinates. For radar point clouds is typically assumed that $\textrm{L}=5$, adding Doppler and power information to the three spatial dimensions, while for lidar point clouds $\textrm{L}=4$ since Doppler is not provided.

\end{itemize}

\subsection{Radar Processing Pipeline Review}\label{radarReview}
The conventional radar processing pipeline is illustrated in Fig.~\ref{RadarPipeline}. The steps are as follows:
\begin{enumerate}
  \item Range and Doppler spectral estimation is performed from the baseband or ADC samples organized in fast-time, slow-time, and channel dimensions. Usually, this is achieved by applying a window with the Fast Fourier Transform (FFT) algorithm independently in fast-time and slow-time. However, this step may be enhanced by compensating the range/Doppler migration due to ego-vehicle and target motion \cite{Xu2022}.
  \item Once a range-Doppler matrix is computed per channel, the angle estimation is performed (1D in azimuth or 2D in both azimuth and elevation, depending on the antenna array topology). Direction of Arrival (DoA) estimation is a current area of widespread interest, with much active research. Usually, Digital Beam-Forming (DBF) is used for simplicity by means of FFT-based implementation, but many research works explore alternatives such as compressive sensing approaches \cite{Rossi2014, Roldan2023TV}, Doppler beam sharpening \cite{Zhang2020, Yuan2024}, or machine learning \cite{Fuchs2022}.  Sometimes, especially in real-time embedded systems, the detection stage is performed before the angle estimation to reduce the computational load \cite{Bilik2018Embedded}, sacrificing the increase in signal-to-noise ratio (SNR) due to spatial coherent integration prior to detection. This process outputs a 4D radar cube.
  \item The detection stage then identifies the cells that contain the targets. Usually, a combination of a CFAR detector in some dimensions and peak finding in the rest is used, though some works have also explored using machine learning algorithms \cite{Brodeski2019,
  Cheng2022, Lin2023,SCORP,kradar,Gusland2020}. In this stage, the data is often sparse since most of the space in the field of view does not reflect sufficient power or is simply empty. The detector outputs a 3D occupancy grid, but a conversion to point cloud is usually performed since it is a convenient format for visualizations or for dataset storage.
  \item After the detection process and the generation of a point cloud, additional steps can be implemented to extract more task-relevant information. For instance, in the automotive context, it is critical to know the nature of each of the detected points to make the appropriate decisions, meaning if this originated from a pedestrian, a vehicle, or some road infrastructure, amongst others. Therefore, it is common to apply a classifier on the point cloud, usually based on DL techniques \cite{Palffy2020, Schumann2020, CRUW}.
  
  If needed for the application, tracking algorithms can also be applied on the point cloud by using past information to reduce the estimation noise, eliminate false detections, and predict future target positions based on the trajectory. In the automotive radar domain, tracking algorithms have to deal with the problem of the extended nature of targets over the angular domain \cite{Kellner2016,Mujtaba2023}.
\end{enumerate}

\section{Related Work}\label{relatedWork}
This work introduces two contributions: the recording \& sharing of the \textit{RaDelft} dataset and the proposed data-driven detector. Therefore, two related work subsections are included to review the state-of-the-art and highlight the need for new radar datasets and new detection algorithms.
\subsection{Radar Datasets}
Several automotive radar datasets have recently been published for different tasks, covering many of the processing steps listed in Section \ref{radarReview}. However, most of them are unsuitable or, at the very least, limited for radar practitioners since the data is already processed, often to the point cloud level. Thus, it is impossible to apply signal processing algorithms that operate on lower level data. Some datasets also provide the radar cube data, but few give the raw ADC data needed to test advanced signal processing methods. 
Essentially, each already-performed processing step limits the scope of the research that can be performed with that data. On the other hand, this simplifies the steps needed to make it suitable for other subsequent tasks. 

A recent summary of the available automotive radar datasets can be found in \cite{Zhou2022}. Nevertheless, in our paper, only those datasets providing data before the point cloud level of processing are considered, since they are the most useful for radar practitioners. Table \ref{tableDatasets} summarizes such datasets.
As can be seen, most of these available datasets are recorded with automotive radars with linear antenna arrays, meaning that there is only azimuth resolution, and no information about the elevation of targets. While useful for some tasks, this type of data is not representative of the data of next-generation 4D radars that are becoming the standard in the automotive field. On the other hand, some datasets already include a 4D imaging radar \cite{radial,colorad,kradar}. The RADial \cite{radial} dataset provides ADC data level suitable for radar practitioners, but the array topology used is not public, and thus, advanced array processing methods cannot be applied. The ColorRadar \cite{colorad} dataset uses a commercially available radar, therefore its datasheet is public. However, most of the scenes are recorded indoors and without a vehicle. Moreover, camera information is not provided. Finally, the K-Radar \cite{kradar} dataset is the most complete, providing range-azimuth-elevation-Doppler cubes, many auxiliary sensors, and useful code to parse the data. However, no ADC-level data is provided, which may limit the potential research scope of the dataset.

Considering the limitations of the aforementioned public datasets, this work presents a new dataset, \textit{RaDelft}, aiming to close the gaps in the existing available datasets collected with a commercially available radar. Our dataset contains three different levels of data processing, namely ADC-level, radar cubes, and point clouds as defined in the previous sections, such that it can serve different future research directions. Additionally, the dataset contains synchronized data from camera, lidar, and odometry, recorded in real-world driving scenarios in the city of Delft. Additional details are provided in the following Section \ref{RaDelftSection}, specifically the sensors used and the developed radar signal processing pipeline.

\begin{table*}[]
\caption{Available public datasets providing either ADC or pre-detection data. In the \textit{DATA TYPE} column, R, D, A, E, and C stand for Range, Doppler, Azimuth, Elevation, and Channel, respectively, while PC means point cloud. In the \textit{ARRAY TYPE} column, 'dense' means that all the half-wavelength spacing is filled with virtual elements. In the \textit{OTHER SENSORS} column, C, L, and O stand for Camera, Lidar, and Odometry, respectively.}
\label{tableDatasets}
\begin{tabular}{lllllll}
\hline
\textbf{Name}   & \textbf{Data Type} & \textbf{Array Type}                                                    & \textbf{\begin{tabular}[c]{@{}l@{}}Virtual Aperture\\ (x$\times$z $\frac{\lambda}{2}$ spacing)\end{tabular}} & \textbf{\begin{tabular}[c]{@{}l@{}}Other \\ Sensors \end{tabular}} & \textbf{\begin{tabular}[c]{@{}l@{}}Record Time \\ (Radar Frames)\end{tabular}} & \textbf{Potential Gaps}                                                                                   \\ \hline
Zendar\cite{Zendar}         & RDC / PC           & Dense ULA                                                              & 4x1                                                                               & CLO                    & 478s (4780)                                                                    & \begin{tabular}[c]{@{}l@{}}No elevation, no ADC data,\\ small aperture.\end{tabular}            \\ \\[-0.5em]
Radiate \cite{Radiate}         & RA                 & \begin{tabular}[c]{@{}l@{}}No array\\ mechanical scanning\end{tabular} & N/A                                                                               & CLO                    & 5h (44000)                                                                             & \begin{tabular}[c]{@{}l@{}}No elevation, no Doppler,\\ no ADC data\end{tabular}                 \\ \\[-0.5em]
CARRADA \cite{Carrada}        & RA / RD            & Dense ULA                                                              & 8x1                                                                               & C                      & 12.1m (12666)                                                                  & \begin{tabular}[c]{@{}l@{}}No elevation, no ADC data,\\ small aperture\end{tabular}             \\ \\[-0.5em]
RADet \cite{Radet}          & RAD                & Dense ULA                                                              & 8x1                                                                               & C                      & 1015s (10158)                                                                  & \begin{tabular}[c]{@{}l@{}}No elevation, no ADC data,\\ small aperture\end{tabular}             \\ \\[-0.5em]
CRUW  \cite{CRUW}          & RA                 & Dense ULA                                                              & 8x1                                                                               & C                      & 3.5h (400000)                                                                    & \begin{tabular}[c]{@{}l@{}}No elevation, no Doppler,\\ no ADC data, small aperture\end{tabular} \\ \\[-0.5em]
Radical \cite{Radical}        & ADC                & Dense ULA                                                              & 8x1                                                                               & C                      & 104m (189000)                                                                    & No elevation, small aperture                                                                    \\ \\[-0.5em]
SCORP  \cite{SCORP}         & ADC / RAD          & Dense ULA                                                              & 8x1                                                                               & C                      & N/A (3913)                                                                         & No elevation, small aperture                                                                    \\ \\[-0.5em]
ColoRadar \cite{colorad}      & ADC / RAE / PC     & Sparse URA                                                             & 86x7                                                                              & LO                     & 145m (43000)                                                                   & No camera, mostly indoor                                                                        \\ \\[-0.5em]
RADial  \cite{radial}        & ADC / RAD / PC     & N/A                                                                      & N/A                                                                               & CLO                    & 2h (25000)                                                                     & Unknow array topology                                                                           \\ \\[-0.5em]
K-Radar  \cite{kradar}       & RAED               & NUA                                                                    & N/A                                                                               & CLO                    & N/A (35000)                                                                          & No ADC data                                                                                     \\ \\[-0.5em]
\textbf{RaDelft (Ours)} & ADC / RAED / PC    & Sparse URA                                                             & 86x7                                                                              & CLO                    & 35m (16975)                                                                    &                                                                                                 \\ \hline

\end{tabular}
\end{table*}

\subsection{Radar Detectors}
The radar detection problem can be formulated as a binary decision task for each radar cube cell, whose objective is to determine whether there is a target or only noise in that specific cell. As mentioned in the previous sections, the automotive radar field has particular challenges when tackling the detection problem. First, the definition of clutter is not univocal in this application since targets of very different natures should be detected, including pedestrians, vehicles, bridges, potholes, road debris, and buildings, among others. Second, since modern automotive radars have high resolution in range, Doppler, and to some extent angle, targets occupy more than a single cell, behaving as extended targets. Finally, not only do the sizes of targets to be detected have a large variance, but also, for the same target, its perceived size can change over time. This is due to two different physical phenomena: the dependency of the angle estimation with its cosine with respect to the radar line of sight, and the relationship of the Cartesian size of the cell with the range due to the angle. Due to all these reasons, conventional CFAR detectors are expected to perform poorly in automotive radar data \cite{Cheng2022, Yoon2019}.

In the past years, several works have been published on detecting extended targets in radar data. Image-based detector techniques have been explored in the literature \cite{Yang2023}, but usually rely on high-contrast data where sharp transitions occur between noise and target. However, due to the finite length nature of signals, spectral leakage in the Fourier processing makes, in general, these transitions soft. Moreover, subspace detectors for extended targets in range and Doppler have been developed \cite{Guan2011, Carretero2011}, but still, an expected spread size of the target energy is needed, in addition to a high computational cost, making them unsuitable for real-time imaging automotive radars.

Also, DL techniques have been applied to the radar detection problem \cite{Brodeski2019, Cheng2022, Lin2023, SCORP,kradar,Zheng2023}. In \cite{Brodeski2019}, a DL detector is proposed, outperforming several 2D CA-CFAR detectors, but only tested in simulated data. In \cite{Gao_2021} and \cite{Lin2023}, the authors propose a similar network structure using 3 autoencoders in three 2D projections (range-angle, range-Doppler, and angle-Doppler) using the annotated dataset in \cite{CRUW} by a camera, avoiding full 3D detection. However, using camera detections as ground truth may be limited due to the 2D nature of camera images. Also, in \cite{Zheng2023}, the authors propose a DL-based detector using bird-eye view radar data, but focusing only on vehicle detection.
On the other hand, the works \cite{kradar} and \cite{Cheng2022} propose two different NNs, but both use the lidar point cloud as ground truth. Since lidar provides high-resolution 3D point clouds, it seems a more reasonable choice to serve as ground truth. The proposed method in \cite{Cheng2022} uses a neural network to detect targets only in the range-Doppler dimensions, followed by the angle estimation and a spatio-temporal filter to enhance the resulting point cloud. On the other hand, in \cite{kradar}, a novel sparse approach to use an NN to detect in the range-azimuth-elevation space is presented. However, the Doppler information is collapsed into a single value, preventing the network from learning the possible angular estimation enhancement due to its relationship with Doppler \cite{Zhang2020, Yuan2024}. Moreover, only the top 10\% power cells are used as input to the network, and therefore, a pre-detection step is used, which can potentially remove target cells. This may be critical in automotive scenarios, where the angular sidelobes of close-range targets may be even 20dB higher than weakly-reflecting distant targets such as pedestrians.

\section{RaDelft Dataset}\label{RaDelftSection}
\begin{figure*}[]
\centerline{\includegraphics[width=\linewidth]{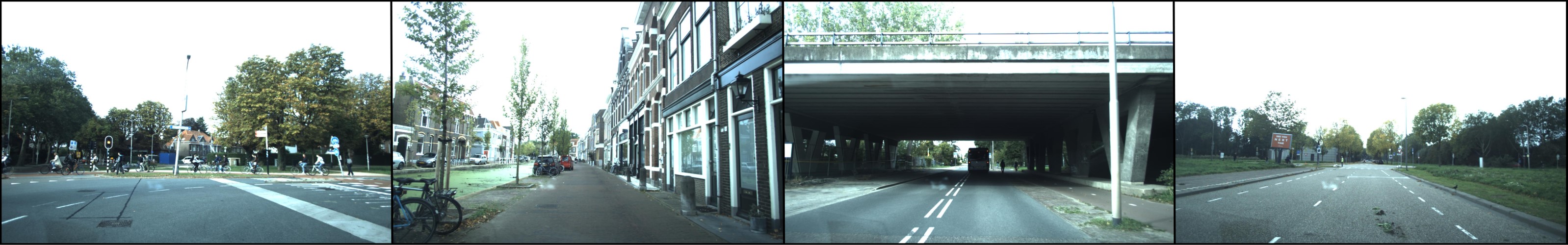}}
\caption{Different frames of different scenes of the \textit{RaDelft} dataset. As it can be seen, there are city center environments, suburban, and different road infrastructures such as large bridges.}
\label{CameraScenes}
\end{figure*}

\begin{figure}[]
\centerline{\includegraphics[width=\linewidth]{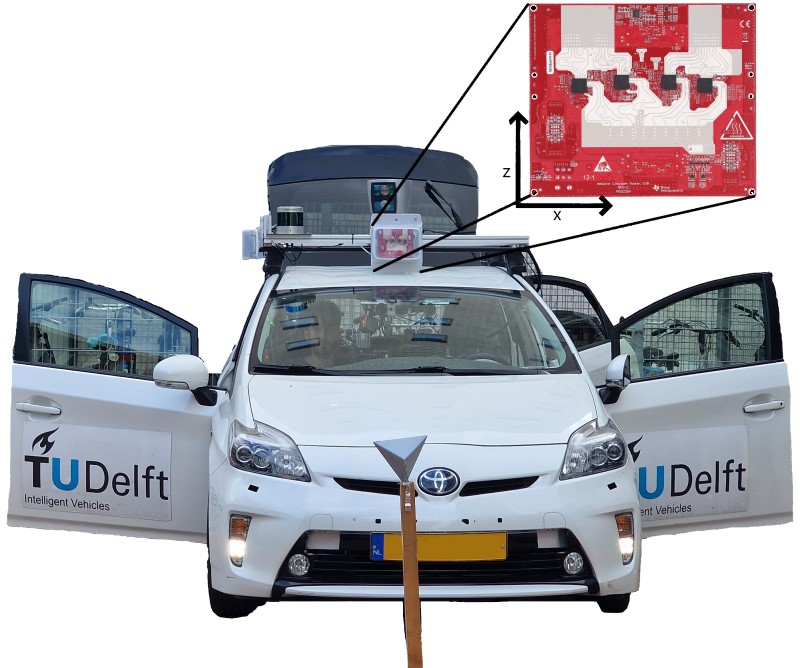}}
\caption{Vehicle used to collect the dataset presented in this paper, equipped with a high-resolution radar, lidar, camera, and odometry. The radar is shown in the top-right inset, with the defined X and Z coordinate axes assumed in this work.}
\label{carFigure}
\end{figure}

The dataset was recorded with the demonstrator vehicle presented in \cite{apalffy2022} with an additional Texas Instrument MMWCAS-RF-EVM \cite{Tidep} imaging radar mounted on the roof at 1.5 meters from the ground. The details of the radar and the waveform used are provided in Section \ref{radarSpecs}. The collection was performed driving in multiple real-life scenarios in the city of Delft with different scene characteristics, such as suburban, university campus, and Delft old-town locations. Four different camera frames are shown in Fig.~\ref{CameraScenes} to illustrate the differences in the environments. 
The output of the following sensors was recorded: a RoboSense Ruby Plus Lidar (128 layers rotating lidar, 10Hz) and the imaging radar board installed on the roof, a video camera (1936 × 1216 px, $\sim$30 Hz) mounted behind the windshield, and the ego vehicle’s odometry (filtered combination of RTK GPS, IMU, and wheel odometry,  $\sim$100 Hz). The sensor setup can be seen in Fig.~\ref{carFigure}. 
All sensors were jointly calibrated following \cite{Domhof2021} and time synchronized. With a 10 Hz frame rate, each scene contains around 2500 radar frames, adding to a total of 16975 frames.

Example code for loading and visualizing the data is provided in a repository\footnote{https://github.com/RaDelft/RaDelft-Dataset} to facilitate the use of the dataset, which can be downloaded from \cite{RaDelft}. Moreover, the radar data is specifically provided at different processing stages for researchers with different backgrounds and interests, including ADC data, radar cubes, and point clouds. The details of the radar processing applied to the data can be found in the next subsection.

\subsection{Radar Configuration and Processing}\label{radarSpecs}
In terms of the specific details of the radar system, this is the MIMO FMCW evaluation board MMWCAS-RF-EVM from Texas Instruments, with 12 transmitters and 16 receivers \cite{Tidep}. The resulting virtual array is an 86-dense uniform linear array (ULA) in the X-direction (shown in Fig. \ref{carFigure}) with half-wavelength spacing, allowing azimuth estimation without grating lobes and a theoretical resolution of $1.33^\circ$ looking at boresight. However, from the point of view of the 2D angular estimation problem in both azimuth \& elevation, the resulting uniform rectangular array (URA) is very sparse, with only a few minimum redundancy arrays (MRA) in the Z-direction (shown in Fig. \ref{carFigure}). Thus, the elevation estimation is very poor in terms of both resolution and ambiguity. The details of the array topology can be found in \cite{Tidep} with graphical representations of the positions of all the elements. Moreover, some elements are overlapped, which can be used to address some of the problems introduced by using time division multiple access (TDMA) in transmission, as detailed later in this sub-section.

\begin{table}
\caption{Radar Waveform Parameters used in the Data Collection\label{waveformParameters}}
\centering
\begin{tabular}{ @{}lr @{}}
\toprule
\textbf{Waveform Parameters} & \textbf{Value}\\
\midrule
Start Frequency (GHz) & 76\\
Effective Bandwidth (MHz) &  750\\
Chirp Slope (MHz/$\mu$s) &  35\\
Chirps Length ($\mu$s) & 28 \\

Idle Time ($\mu$s) & 5 \\
Number ADC Samples per Chirp & 256 \\
Number of Chirps per Frame& 128 \\
Sampling Frequency (Msps) & 12 \\
Tx strategy  & TDMA \\
\toprule
\textbf{Derived Quantities} &  \textbf{Value}\\
\midrule
Range Resolution (m) & 0.2 \\
Maximum Unambiguous Range (m) & 51.4 \\
Velocity Resolution (m/s) & 0.046 \\
Maximum Unambiguous Velocity (without extension) (m/s) & 2.48 \\
Maximum Unambiguous Velocity (with extension) (m/s) & 17.36 \\
\bottomrule
\end{tabular}
\end{table}

The radar waveform parameters used can be seen in Table \ref{waveformParameters}, with the derived resolution and ambiguity values.
The complex baseband samples are saved in the dataset using the same format provided by the radar manufacturer, but MATLAB code is provided to parse it, reshape it into a $\textrm{N}_{fast} \times \textrm{N}_{slow} \times \textrm{N}_{Vchan}$ 3D tensor, and process it to the radar cubes.

The first step of the processing is to apply a Hamming windowing and an FFT in the fast-time and slow-time dimensions to perform range and Doppler estimation. Then, the detrimental effects of the TDMA have to be compensated. The first effect is related to the extension of the Pulse Repetition Interval (PRI) by a factor equal to the number of transmitters. Therefore, the maximum unambiguous Doppler and the corresponding maximum measurable velocity (without ambiguity) $v_{max}$ is reduced, as can be seen in equation (\ref{vmax}): 
\begin{equation}
    v_{max} = \frac{c}{4f_cPRI},
    \label{vmax}
\end{equation}
where $c$ is the speed of light and $f_c$ is the carrier frequency. 
This effect is especially problematic in the automotive context, where targets can have high relative speeds. Moreover, the phase difference between signals received from different transmitters will depend on both the angle of arrival of the signal and the velocity of the targets, due to the target’s movement between transmission times of different transmitters operating in TDMA mode \cite{Zoeke2015}. 
This resulting phase migration term is shown in equation (\ref{phasemigration}): 
\begin{equation}
    \phi_{mig} = \frac{4\pi}{\lambda}v\Delta t,
    \label{phasemigration}
\end{equation}
where $\lambda$ is the wavelength, $v$ is the relative speed of the target, and $\Delta$t is the time difference between transmitters. This term must be compensated before performing angle estimation to avoid significant artefacts. 

In this work, both undesirable effects of TDMA are solved by using the overlapped virtual antennas present in the radar system with the algorithms provided in \cite{Schmid2012}. However, it is important to take into account that the maximum unambiguous velocity extension only works when a single target is present in a range-Doppler cell. Therefore, if multiple targets are folded into the same Doppler bin, or there are targets in different angles at the same range-Doppler bin, the algorithm will not be able address the problem. 
Since this work does not aim to solve the Doppler ambiguity problem in TDMA, the aforementioned constraint is accepted as a limitation of the current commercial radar system. 
Nevertheless, it is assumed that making the ADC samples directly available in our dataset can be valuable for the research community, for example to apply more advanced approaches for Doppler/velocity ambiguity in TDMA in the future.

The angle estimation can be performed once the TDMA effects have been compensated. It is important to remember that the resulting virtual array is a very sparse URA with some structures. While other research works deal with this type of array, for instance by trying to fill/interpolate the missing elements or applying compressive sensing techniques \cite{Shunqiao2021, Rossi2014}, the core of this work is not to improve the angular estimation with sparse arrays. Therefore, a very simple approach of zero-filling and FFT processing has been applied. However, due to the sparseness of the radar antenna array in the Z-direction (shown in Fig. \ref{carFigure}), grating lobes and high-side lobes appear in elevation. To mitigate this problem, the field of view in elevation has been restricted to $\pm15^\circ$ degrees, and the elevation value with the highest power has been selected and saved, discarding the rest.
Also, the azimuth estimation has been restricted to $\pm70^\circ$ for two reasons. First, the angular estimation performance outside this region is rather poor, as:
\begin{equation}
    \Delta\theta \sim \frac{1}{\cos{\theta}},
    \label{angleResolution}
\end{equation}
being $\Delta\theta$ the angular resolution and $\theta$ the estimated angle. Secondly, the radiation power is almost 10dB lower than at boresight outside this region, making target detection very challenging. 

Subsequently, after zero-padding, FFT processing and field of view (FoV) cropping, the resulting radar cubes have dimensions $\textrm{N}_{r}\times \textrm{N}_{D} \times \textrm{N}_{a} \times2$ $ (500 \times 128 \times 240 \times 2)$. This essentially means that for each range-Doppler-azimuth cell, there are two values: the elevation value with the highest detected power level, and the power level itself. 
Note that the $240$ azimuth bins span the $\pm70^\circ$ degrees of the FoV after cropping, but not uniformly, due to the non-linear relation in the equation (\ref{angleResolution}). 
For simplicity and to save storage space in the shared dataset, the aforementioned values are saved as different cubes since the elevation can be stored as an integer number (i.e., denoted as elevation bin), while the power value is a float.

Finally, a detection stage is applied to the radar cubes to generate a point cloud. This lower dimensionality representation of the data is also provided within the shared dataset to ease the process for researchers who want to use this highly processed data straightforwardly without going into the details of radar signal processing.

\section{Proposed Data Driven Detector}
\label{datadrivendetector}
\begin{figure*}[t]

\centerline{\includegraphics[width=\linewidth]{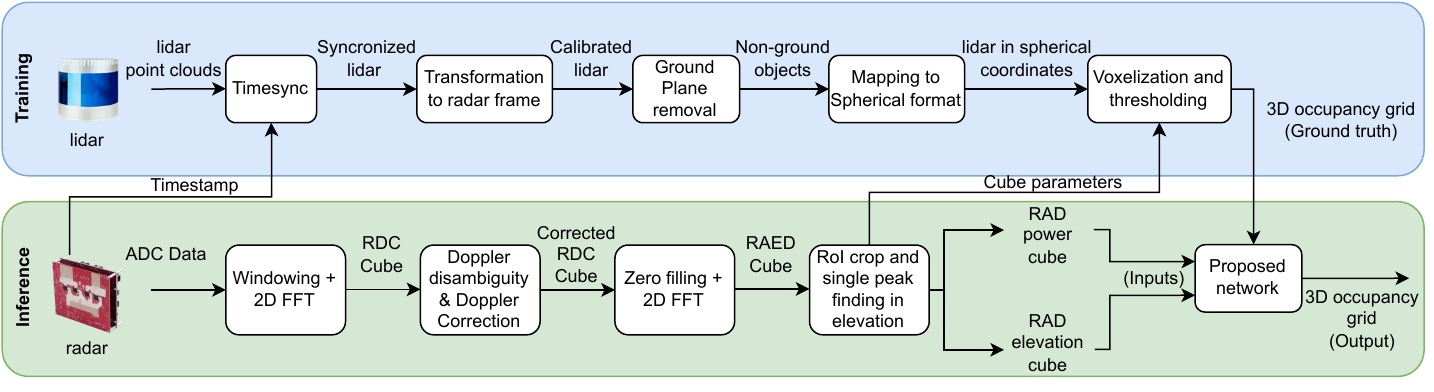}}
\caption{Overview of the proposed data-driven detector. The steps to generate the 3D lidar occupancy grid are shown on the top row, which will be then used as ground truth for training the neural network. The radar signal processing pipeline is shown at the bottom of the figure, and is needed to generate the input data for the network. RDC stands for range-Doppler-channel (no angle estimation), and RAED stands for range-azimuth-elevation-Doppler \cite{roldan2024cfar}. RoI stands for region of interest.}
\label{fig:ProposedMethodFigure}
\end{figure*}

To address the aforementioned shortcomings of current detectors in automotive radar, a novel data-driven detector is proposed to generate 3D occupancy grids only with radar data, using neural networks and lidar data as ground truth. A visual summary of the method can be seen in Fig.~\ref{fig:ProposedMethodFigure}. 

\begin{figure*}[t]
\subfloat[]{\includegraphics[width=.33\textwidth]{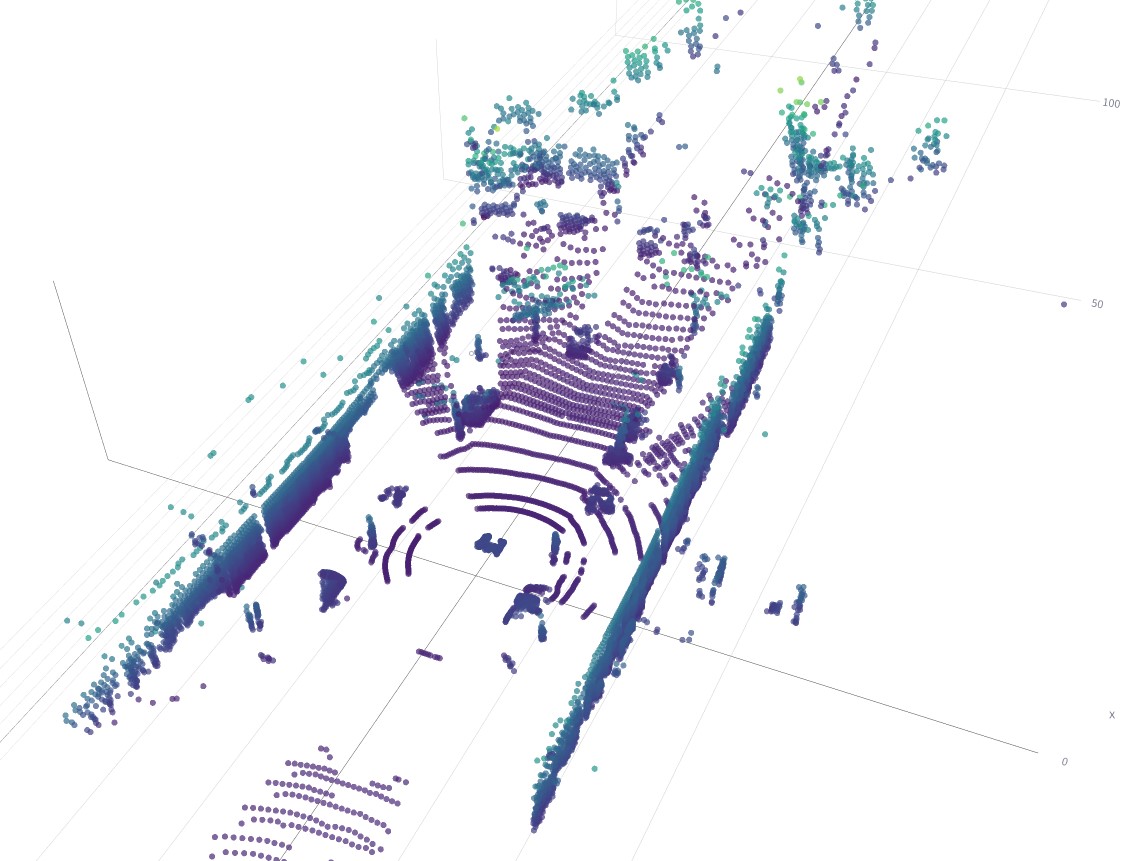}%
\label{LidarPCFull}}
\subfloat[]{\includegraphics[width=.33\textwidth]{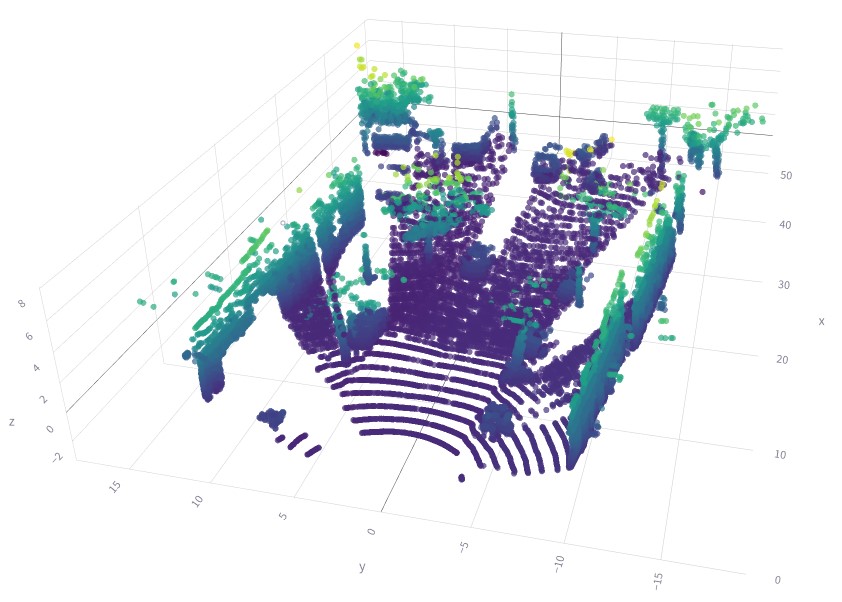}%
\label{LidarPCRoI}}
\subfloat[]{\includegraphics[width=.33\textwidth]{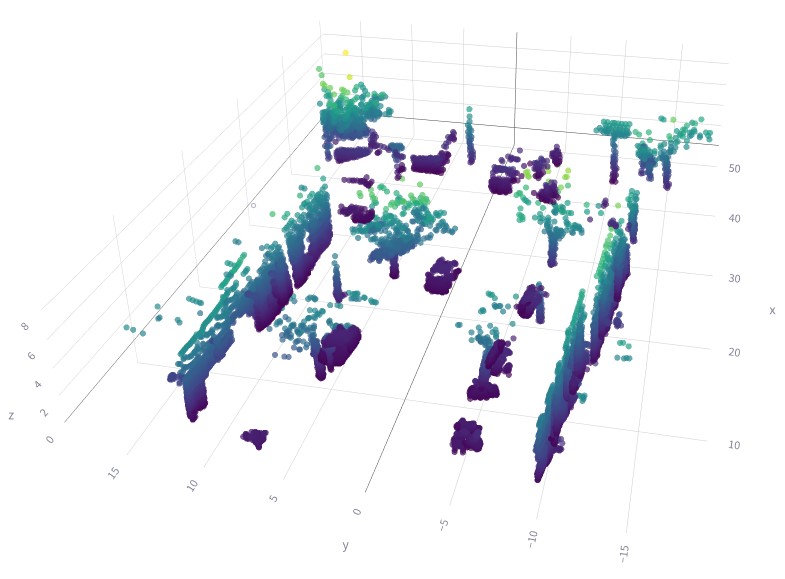}%
\label{LidarPCFinal}}
\caption{In (a) the original point cloud as provided by the lidar sensor. In (b) the lidar point cloud after cropping to mimic the radar field of view (i.e., $\pm 70^{\circ}$ in azimuth and $\pm 20^{\circ}$ in elevation). In (c), the lidar point cloud after the road surface removal using PatchWork++ \cite{Lee2022}, which will be used as ground truth to train the proposed data-driven detector.}
\label{LidarPC}
\end{figure*}

The first step of the method is to adapt the lidar point cloud to serve as the ground truth. For each radar cube, the closest lidar point cloud in time is selected based on the timestamps for both radar and lidar data, assuming that a small error due to different start times may be present. Since the lidar system used in this work is mechanically rotating, it provides $360^\circ$ coverage. Therefore, the first step is to crop this as to the same FoV of the radar, i.e., $\pm20^\circ$ in elevation \& $\pm70^\circ$ in azimuth, and a maximum range of 50m. To illustrate this difference in the FoV, Fig.~\ref{LidarPCRoI} shows the cropped lidar point cloud compared to the original point cloud in Fig.~\ref{LidarPCFull}. Moreover, removing all the lidar points from the road surface is essential as the road surface is hardly visible to the radars and could lead to noisy ground truth for the training process. The Patchwork++ algorithm is used to this end \cite{Lee2022}. 
After removing the road surface points, the resulting lidar point cloud can be seen in Fig.~\ref{LidarPCFinal}. 

Finally, the processed lidar point cloud has to be converted into a 3D cube to serve as ground truth. This voxelization process can be understood as generating a 3D occupancy grid, where each voxel contains ‘one’ if at least one lidar point is inside, and ‘zero’ otherwise. However, it is important to note that the radar cube grid is not uniform due to the Fourier Transform processing for angular estimation and its relationship with the cosine of the estimated angle. This effect, which essentially makes the cells thinner at boresight and broader at the edge of the field of view, must be considered to generate the same non-uniform lidar 3D occupancy grid. It is important to notice that all this process can be performed offline, outside the NN training loop, saving the processed lidar point clouds beforehand to speed up the training.

\begin{figure}[]
\centerline{\includegraphics[width=\linewidth]{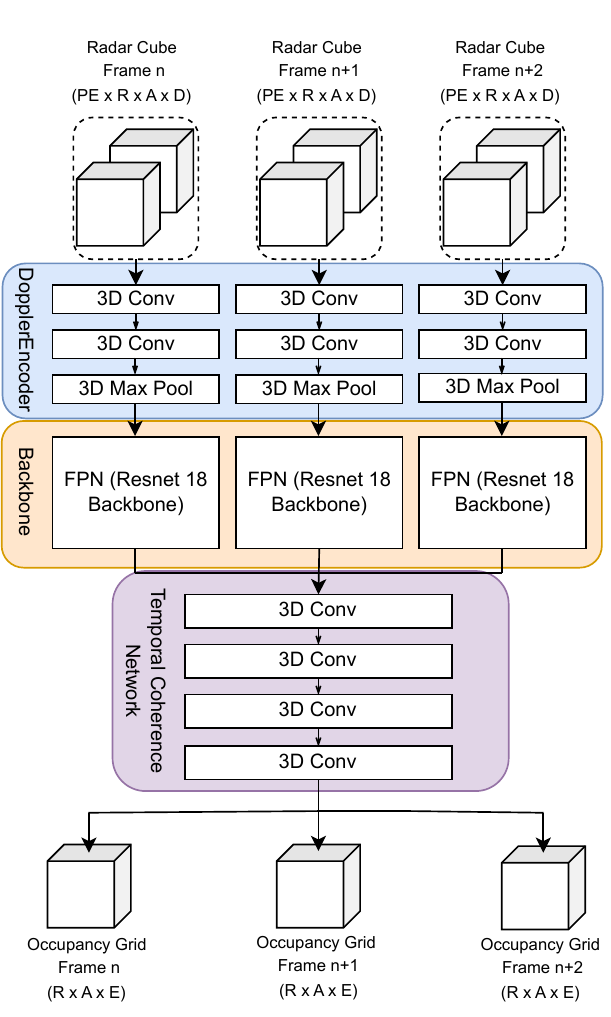}}
\caption{Proposed network architecture for the data-driven detector composed of three sub-networks. First, the \textit{DopplerEncoder} network aims to encode the Doppler information so that its information is retained even if not directly comparable with ground truth lidar data. Then, a standard FPN with a Resnet \textit{backbone} is used. Note that the three branches process separately three frames of data but share the same weights. Finally, the three outputs are concatenated to produce an input tensor to the \textit{temporal coherence network} which generates the final occupancy grid for each frame.}
\label{Network}
\end{figure}

Once the ground truth has been appropriately generated as described above, the NN can be trained. The proposed NN is an evolution of the previous model validated in \cite{roldan2024cfar}. Specifically, in this case, the network is modified to use three frames of data as input to model temporal patterns, and the NN predicts the 3D occupancy grid for the three frames simultaneously. 
This modification has been implemented to reduce the 'flickering' usually present in the radar point clouds, where isolated points pass the detection threshold due to instantaneous high noise but disappear in consecutive frames. Therefore, the proposed NN tries to enforce some temporal consistency. 
A diagram of the complete network architecture is shown in Fig.~\ref{Network}. As it can be seen, the input is a $\mathrm{T} \times 2 \times \mathrm{R} \times \mathrm{A} \times \mathrm{D}$ tensor, where in practice, $\mathrm{T}=3$ (frames), $\mathrm{R}=500$ (range bins), $\mathrm{A}=240$ (azimuth bins), and $\mathrm{D}=128$ (Doppler bins). As explained in the previous section, these values are higher than the initial number of fast-time samples, slow-time samples, and virtual channels due to zero padding applied before the FFT processing. Moreover, the number of frames $\mathrm{T}=3$ has been chosen as a trade-off between managing to capture temporal information and losing useful correlation between frames since the scene is often not static, and including too many frames will result in inconsistencies.

In terms of architecture, the first part of the proposed NN is the \textit{DopplerEncoder} subnetwork. As the lidar cannot measure Doppler information, the detections on the Doppler dimension of the radar data cannot be directly utilised and compared to the ground truth. However, there is a known relationship between Doppler and angle in the case of moving platforms (or moving targets). Thus, the Doppler dimension is not simply removed from the radar data but rather encoded so that it can still be used in the overall detection process, as it may be beneficial for angular estimation.
Specifically, here the \textit{DopplerEncoder} subnetwork extracts all the Doppler information in each range-azimuth cell and encodes it into the channel dimension. 
This is achieved by using two 3D convolutional layers followed by a 3D max pool layer, transforming the $2 \times \mathrm{R} \times \mathrm{A} \times \mathrm{D}$ input tensor into a $64 \times \mathrm{R} \times \mathrm{A}$ tensor, where the 64 channel dimension contains the encoded information of Doppler and elevation. 

The second part of the proposed network is an off-the-shelf 2D CNN \textit{backbone}, applied to estimate the final $\mathrm{R}\times \mathrm{A} \times \mathrm{E} \ (500\times 240 \times44)$ 3D occupancy grid. The significant advantage of using such 2D CNN backbones is their compatibility with hardware accelerators (e.g., GPUs and TPUs) and major machine learning frameworks (e.g., Tensor-Flow, PyTorch), leading to enhanced computational efficiency. 
While the current proposed implementation employs a Feature Pyramidal Network (FPN) \cite{FPN} with a Resnet18 backbone \cite{he2015deep}, our modular design allows for different architectures to be used for this purpose, enabling the system to be tailored to the specific memory and computational requirements of the intended platform. 

These two parts of the proposed network are applied to each of the three considered frames independently, as shown in Fig. \ref{Network}, but the weights of the layers are shared, and the output is concatenated into a $\mathrm{T} \times \mathrm{R} \times \mathrm{A} \times \mathrm{E} \ (3 \times 500 \times 240 \times 44)$ tensor. 
Finally, to take into account the temporal relationship between the three frames, a third module composed of six 3D convolutional layers is included (referred to as \textit{TemporalCoherenceNetwork} in Fig. \ref{Network}). It is important to notice that even if the output is a 3D occupancy grid for each frame, the power information on each cell is not lost, since the indices of the detected cells from such grid can be used to retrieve the corresponding intensity information from the original RAED cube.

One of the key characteristics of the radar data is the scene sparsity. Of all the voxels in the generated 3D occupancy grid, only around 1\% contain targets. Therefore, this must be considered when selecting the loss function for training the neural network. In this work, the Focal loss \cite{lin2018focal} is used for this purpose, which handles class imbalances in a similar way to the weighted cross-entropy loss, and adds an extra modulating factor to focus on the hard cases. The Focal loss \cite{lin2018focal} is defined as:
\begin{equation}
    FL(p_t) = -\alpha_t(1-p_t)^\gamma \log (p_t),
    \label{focalLoss}
\end{equation}
with 
\begin{equation}
    p_t = \begin{cases}
       p & \textrm{if} \quad  y=1\\
    1-p & \textrm{otherwise}\\
     \end{cases}
    \label{focalLoss}
\end{equation}
where $y \in \{\pm 1\}$ is the ground-truth class (i.e., detection or not), $\alpha_t$ the weighting factor to take into account data imbalance defined as $\alpha \in [0,1]$ for class 1 and $1 - \alpha$ for class -1, and $\gamma > 1$ is the focusing factor. This loss is especially interesting in radar data, since high RCS targets can be easily detected, but low RCS targets or targets located at a far distance are more challenging to detect, and this can be taken into account by the $\gamma$ parameter.
In terms of training-testing split, 90\% of the data from five of the seven recorded scenarios have been used to train the network using Adam optimizer, leaving 10\% for validation. The network was trained using the DelftBlue Supercomputer \cite{DHPC2024} from TU Delft. The remaining two recorded scenarios are used as a test set, i.e., with data completely new, unseen for the network.

\section{Results}\label{resultsSection}
The trained neural network can estimate the 3D occupancy grid for each radar cube and thus act as a detector. It should be noted that all the results presented in this section are evaluated using only the test set, composed of the 2 scenes left out from the training process. This ensures data independence and, while still collected in the same geographical area, the capability of the proposed method to generalize to unseen data with different characteristics. 

Two main performance metrics are used to evaluate the results of the proposed neural network: the usual probability of detection (P$_d$) and probability of false alarm (P$_{fa}$) metrics, and the Chamfer distance (CD). In both cases, the lidar data is used as a reference, either in the occupancy grid format for the P$_{d}$ and P$_{fa}$ computation, or in the point cloud format for the Chamfer distance. While different definitions are given for the Chamfer distance in the literature, in this work the following is used:
\begin{equation}
\begin{split}
    CD(S_1,S_2) = \frac{1}{|S_1|} \sum_{x\in S_1} \min_{y\in S_2} ||x-y||_2 +  \\ \frac{1}{|S_2|} \sum_{y\in S_2} \min_{x\in S_1} ||x-y||_2,
    \label{chamfer}
\end{split}
\end{equation}
where $S_1$ and $S_2$ are the two sets of points being compared (e.g., the lidar points assumed as ground truth vs the points from the 3D occupancy grid provided by the proposed data-driven detector), and $|S|$ is the cardinality of the set. The closer the two sets of points are the better, and so the lower the Chamfer distance.

\begin{figure}
\centerline{\includegraphics[width=\linewidth]{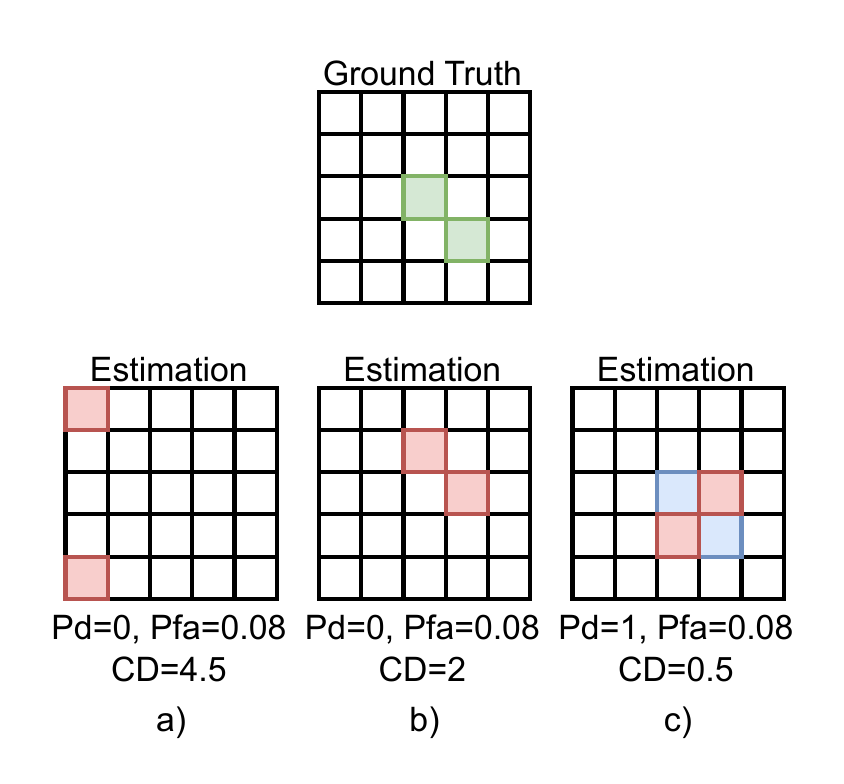}}
\caption{Illustration of the problem in computing the P$_d$, P$_{fa}$, and Chamfer Distance (CD) as performance metrics. In a), a case where two ghost targets are created. In b), a calibration misalignment shifts the detection cells, raising the P$_{fa}$ as if two ghost targets were created. In c), the problem of the overestimation of target size. These three cases have nominally the same P$_{fa}$, but the implications for overall scene perception are completely different. It can be seen how the CD captures the spatial relationships and yields a better value in cases b) and c), where the false alarms have less impact from an application point of view. }
\label{pdpfa}
\end{figure}

However, it is important to note that a caveat is needed when analyzing the P$_d$ and the P$_{fa}$ metrics. A small misalignment in the calibration of only a few centimeters in range or of a small angle will cause the probability of detection to fall drastically, while the probability of false alarms will rise, as can be seen in the examples presented in Fig.~\ref{pdpfa}. For instance, considering the example in Fig.~\ref{pdpfa}b, even though the $P_{fa}$ of this case is numerically the same as in the case represented in Fig.~\ref{pdpfa}a, the impact in terms of quality of the perceived environment can be very different, especially taking into account the small cell dimensions. While this is not a problem in the proposed method (as the network used as data-driven detector is able to learn offsets such as those in this example), it may affect the other methods used in this section for benchmarking, such as the different variants of CFAR detectors. 
Moreover, since the radar resolutions are worse than the lidar's, many targets will be overestimated in size, raising the $P_{fa}$. These false alarms are, in general, assumed to be less relevant for assessing the quality of automotive radar since a small overestimation of objects in the order of centimeters (i.e., few lidar resolution cells) may not be as bad as detecting isolated ghost targets. Nevertheless, all the false alarms are treated equally in the assessment in this paper, since an extra clustering or tracking stage may be needed to distinguish between these unfavourable cases in terms of $P_{fa}$. An example of this phenomenon can be seen in Fig.~\ref{pdpfa}c. On the other hand, it can be seen how the Chamfer distance is able to capture these spatial relationships, yielding different values for the three different cases. Taking all this into account, a point cloud level metric like the Chamfer distance is considered to be a better evaluation metric for this work.

Table \ref{tab:tableComprehensiveResults} shows the performance of the proposed method with the three aforementioned metrics averaged over the whole test set and compared with different alternative approaches for detection. Specifically, the different rows on the table are:
\begin{itemize}
  \item \textit{Proposed Method}: the results of the proposed method explained in the previous section and with overall architecture shown in Fig. \ref{Network}.
  \item \textit{No Doppler \& Quantile}: an approach similar to the one presented in \cite{kradar}, where only those power cells with values higher than the 0.9 quantile are kept and the rest is set to zero. Furthermore, the Doppler information is collapsed by taking the mean over the Doppler dimension. This is used to 'sparsify' the data and speed up processing, with the risk of cutting out weakly reflecting targets. Since there is no Doppler data anymore, the \textit{DopplerEncoder} subnetwork is removed from the general architecture of the proposed data-driven detector. It is important to note that the segmentation backbone has 13.2 million parameters, while the \textit{DopplerEncoder} subnetwork only 76.9K. Therefore, the comparison between the full network and the network without the \textit{DopplerEncoder} is possible without adding extra layers.
  \item \textit{Quantile}: the proposed method, but with the pre-detection fixed threshold based on the 0.9 quantile inspired by \cite{kradar}.
  \item \textit{No Time}: in order to assess the impact of inputting several frames into the network and use the temporal evolution of the scene, this tests the proposed method without the \textit{Temporal Coherence} subnetwork in the architecture, essentially an ablation study without inter-frame temporal information.
  \item \textit{OSCFAR}: a 2D OS CFAR in range-angle, followed by a 1D OS CFAR in Doppler. While multiple different CFAR alternatives have been tested (i.e., different combinations of CA and OS CFAR detectors), only the best implementation is reported here for conciseness. An analysis with different variations has been presented in \cite{roldan2024cfar} for completeness. Following \cite{Rohling83}, the rank has been set to 0.75 times the number of training cells, and no guard cells have been used.
  
\end{itemize}

As it can be seen in Table \ref{tab:tableComprehensiveResults}, the highest P$_d$ and the lowest Chamfer distance is achieved by the proposed method while maintaining a similar P$_{fa}$. On the other hand, applying the quantile cut and removing the Doppler information similarly to \cite{kradar} reduces the P$_{d}$ from 62.13\% to 52.97\%, and worsens the Chamfer distance from 1.54m to 2.16m. Looking at the results for the other versions, it can be seen that this drop in performance is mostly due to the removal of the Doppler information. Using only the quantile-based threshold may be a good trade-off since the performance degradation is not substantial, but the computational cost is reduced. Looking at the version without the \textit{Temporal Coherence} subnetwork, which is trained on single frames, it can be seen how all the metrics are worse than in the baseline. Thus, including temporal information in the network is a good strategy to boost performance, with the only downside of increasing slightly the training time due to the extra layers. Finally, it can be seen that the conventional OS CFAR is the method that performs the worst, with a much higher Chamfer distance of 6.73m.

In order to have a fairer comparison against the conventional CFAR detector, a 2D version of the proposed method has also been evaluated by disregarding the elevation information, as this can only be estimated rather poorly due to the unfavorable design of the radar array. To this end, the proposed NN has been trained without elevation information, discarding the virtual channels in the Z-direction and, thus, treating it as a ULA in the azimuth direction. For completeness, the implementation with quantile based threshold has also been assessed in this new analysis. The results are shown in Table \ref{tab:tableComprehensiveResults} under the 'No Elevation cases' rows.
For these tests, the P$_d$ of the OS CFAR approach is increased to 11.5\%, but the P$_{fa}$ is also raised. This is mainly due to detections triggered in the adjacent angle bins of a target generating "ring like" patterns due to side lobes, a phenomenon also mentioned in \cite{kradar}. 
Both the proposed method and the proposed method with the quantile based threshold are shown to outperform the conventional OSCFAR in the three metrics.

\begin{table}

\caption{Performance Results of the proposed method for data-driven detection, different variations of the method, and the best-performing CFAR detector implemented.}

\centering
\begin{tabular}{@{}cccc@{}}
\toprule
\textbf{Method} & \boldmath{$P_d$ (\%)} & \boldmath{$P_{fa}$ (\%)} & \textbf{Chamfer distance ($m$)} \\
\toprule
\textbf{Proposed Method} & 62.13 & 2.77 & 1.54\\
No Doppler \& Quantile & 52.97 & 2.63 & 2.16\\
No Doppler  & 50.44 & 2.50 & 2.13\\
Quantile & 57.9 & 2.85 & 1.92\\
No Time (single frame) & 58.08 & 2.63 & 2.16\\
OSCFAR & 0.41 & 0.015& 6.73\\
\toprule
\textbf{No Elevation cases} &  \\
\midrule
\textbf{Proposed Method} & 74.83 & 1.12 & 2.92\\
Quantile & 74.09 & 1.11 & 2.78\\
OSCFAR & 11.56 & 3.1 & 4.11\\

\bottomrule
\end{tabular}
\label{tab:tableComprehensiveResults}
\end{table}

\begin{figure*}[t]
\centerline{\includegraphics[width=\linewidth]{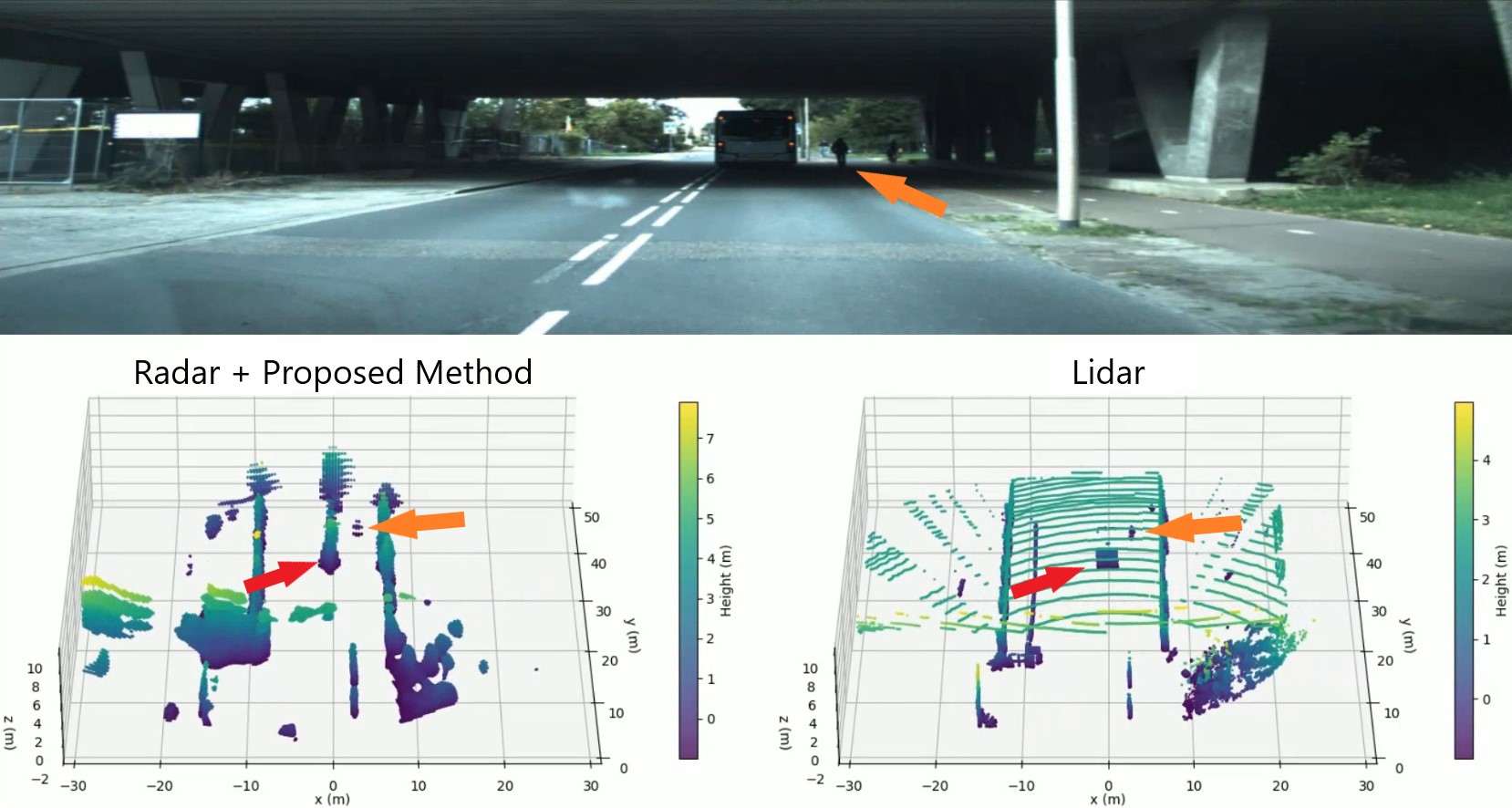}}
\caption{Example frame in a challenging situation for the radar system, where the vehicle is going under a large bridge. In the top figure, the camera image is shown for reference. On the left, the point cloud generated with the proposed data-driven method is shown, and on the right the original point cloud provided by the lidar. The red arrows point to the bus under the bridge and the orange arrow to the pedestrian next to it. Note that the color in the point clouds refer to the height of the objects.}
\label{Example2}
\end{figure*}

In addition to the quantitative results, some qualitative results are also presented to show the performance of the proposed method visually. In Fig.~\ref{Example2}, a challenging frame from the radar point of view is shown, where the vehicle is going under a large but relatively not tall bridge. The 3D point cloud generated with the proposed method is shown in the left plot, with the original lidar on the right plot. As it can be seen, the road is clear of false alarms, and the bus (in red arrow) and pedestrian (in orange arrow) are clearly detected. The bus and the ceiling merge due to the poor elevation resolution of the radar data, but they could be split and identified in Doppler. 

\begin{figure*}[t]
\centering
\subfloat[]{\includegraphics[width=.325\textwidth]{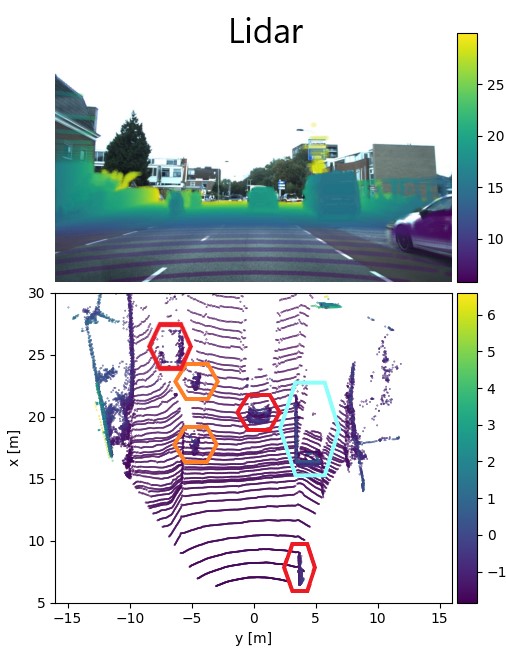}%
\label{resultsLidar}}
\subfloat[]{\includegraphics[width=.325\textwidth]{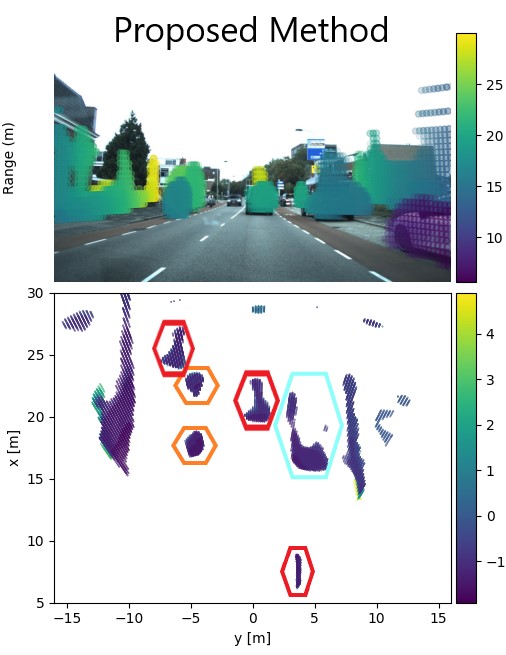}%
\label{resultsNN}}
\subfloat[]{\includegraphics[width=.335\textwidth]{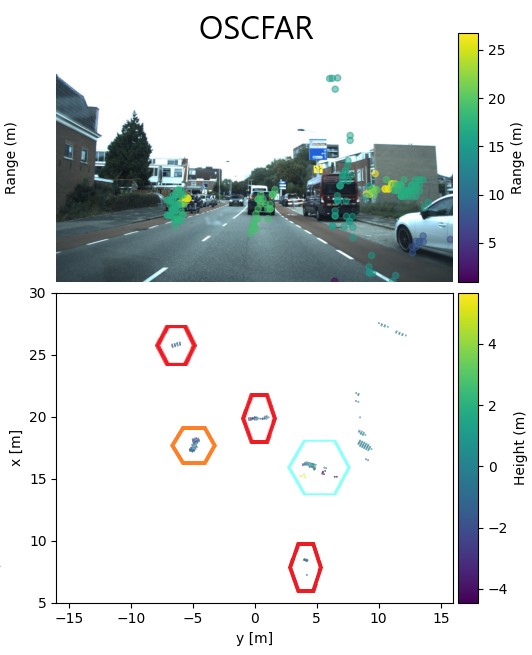}%
\label{resultsCfar}}
\caption{Example of data frame in urban scenario with related detections. In (a), the original lidar point cloud projected onto the camera as well as a bird's eye view. In (b), the radar point cloud generated with the proposed data-driven detector. In (c), the radar point cloud generated with the best-performing CFAR implemented (i.e., 2D OS-CFAR in range-azimuth, followed by an OS-CFAR in Doppler).}
\label{Results}
\end{figure*}

Fig.~\ref{Results} shows another scene where the resulting point clouds have been projected onto the camera image to provide a sense of the 3D scene (top), but the bird's eye view projection is also shown (bottom). For simplicity, the point clouds have been cropped to a maximum range of 30 meters. Moreover, as a visual aid in the bird's eye view, cyclists are highlighted with an orange hexagon, cars with a red hexagon, and a large van with a light blue hexagon.
In Fig.~\ref{resultsLidar}, the original lidar point cloud is presented, where many details of the scene can be appreciated. Fig.~\ref{resultsNN} shows the detections generated using the proposed data-driven detector, and as it can be seen, most of the details of the relevant targets are preserved. Objects are slightly overestimated in size, but the overall scene is clear. Also, the shape of the objects is preserved, especially in the case of cars and the large van. Finally, Fig.~\ref{resultsCfar} shows the output of the previously-mentioned best-performing CFAR detector, where it can be seen how the output is much sparser in terms of detected points, and also missing one of the cyclists in the scene.

\begin{figure*}[t]
\centerline{\includegraphics[width=\linewidth]{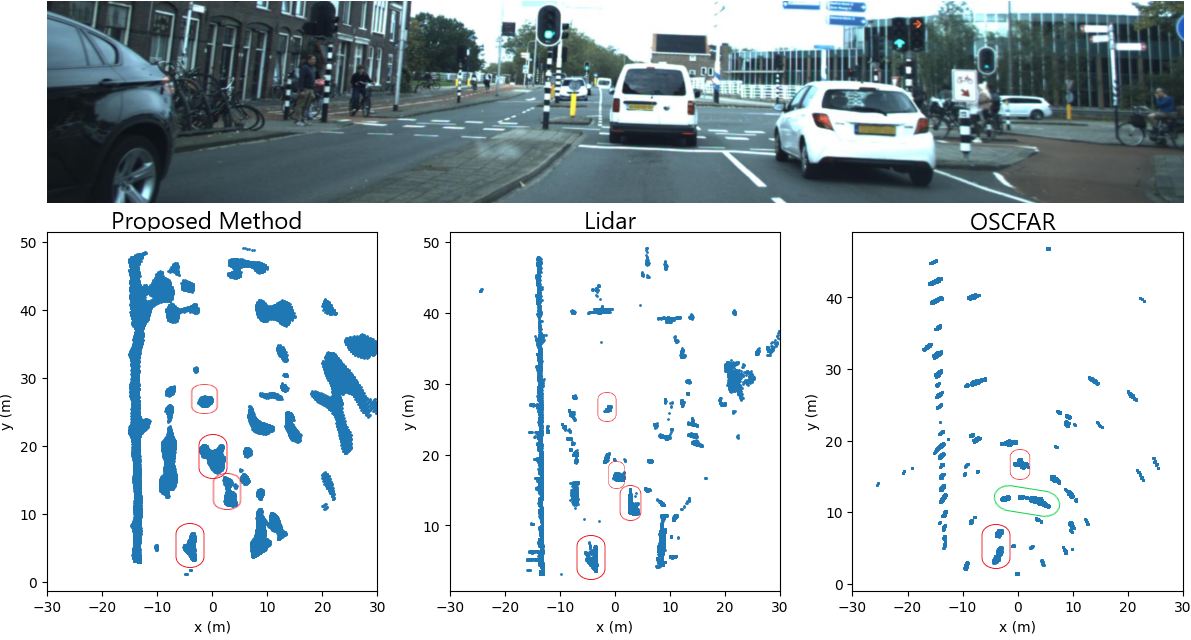}}
\caption{Example of data frame where the elevation information is disregarded from the detection process. In the top figure, the camera image is shown for reference. In the bottom part of the figure, the original lidar point cloud is shown (center), with the point cloud generated by the proposed data-driven detector (left) and by the best-performing implemented CFAR (right). }
\label{bev}
\end{figure*}

Finally, an example of results where the elevation information is disregarded in the detection process is presented in Fig.~\ref{bev}. Here, the figure shows the camera image for visual reference (top), and the comparison of the resulting point cloud from the radar data with the proposed data-driven detector (left), the original lidar data (center), and the point cloud from the radar data with the best-performing implemented CFAR. Note that cars are highlighted in red, and there are "ring-like" detections (highlighted in green) due to the high side lobes of the van, which can be seen in the figure generated using the CFAR detector. This phenomenon raises the P$_{fa}$ and is an expected behavior that has been reported in other automotive radar datasets \cite{kradar} when using CFAR detectors. As also reported in the previous qualitative examples, the point cloud generated by the proposed data-driven detector is denser than the CFAR-generated one, and conserves the correct location and shape of most objects.

\section{Conclusions}\label{conclusionsSection}
This work introduces an innovative data-driven detector for automotive radar and the \textit{RaDelft} dataset, a newly collected multi-sensor real-world dataset. The proposed radar detector is trained exclusively from unlabeled synchronized radar and lidar data, thus eliminating the need for costly manual object annotations for the detection process. Two types of performance metrics were employed to validate the method, i.e., conventional probability of detection \& probability of false alarm, alongside the Chamfer distance, a point cloud-level metric designed to capture spatial relationships and similarities between point clouds. 
The proposed method reduces by 4.2 meters (77\% reduction) the Chamfer distance when compared with conventional OSCFAR detectors, and by 0.62 meters (28\% reduction) when compared with the state-of-the-art. Also, it significantly increases the probability of detection. Moreover, an ablation study showed that including temporal information in the process is important, and Doppler information is especially crucial for our model’s good performance. Results show that the probability of detection is increased from 50.44\% to 62.13\%, and the Chamfer distance is reduced by 27\% when using Doppler information.

For the experimental evaluation of the proposed approach, a comprehensive dataset encompassing over 30 minutes of actual driving scenarios was collected using a vehicle equipped with both lidar and radar sensors, resulting in 16975 radar frames paired with corresponding lidar ground truth. Compared with other existing datasets, \textit{RaDelft} provides raw data from a commercial 4D imaging radar needed for radar practitioners for many research lines. Moreover, it contains data processed at other levels (e.g., radar cubes and point clouds) suitable for researchers with different backgrounds and interests. The dataset is publicly available, with code to parse, visualize, and process the data, as well as the code to reproduce the results reported in this work.

\bibliographystyle{IEEEtran}
\bibliography{refs}


\begin{IEEEbiography}[{\includegraphics[width=1in,height=1.25in,clip,keepaspectratio]{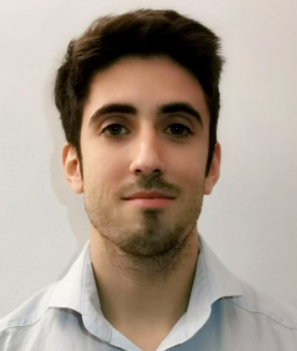}}]{Ignacio Roldan}
received his B.Sc. and M.Sc. in Telecommunication Engineering at the Universidad Politécnica de Madrid, Spain in 2014 and 2016. In 2018 he complemented his education with an M.Sc. in Signal Processing and Machine Learning at the same university. He has worked for more than 5 years at Advanced Radar Technologies, a Spanish tech company focused on the design and manufacture of radar systems. During this period, he has been involved in several international projects developing state-of-the-art signal processing techniques for radars. In his last stage, he was focused on applying Machine Learning techniques to UAV detection and classification. In Sep 2020, he joined the Microwave Sensing, Signals, and Systems group at Delft University of Technology, where he is working towards the PhD degree.

Mr. Roldan received the best student paper award at the 2024 IEEE Radar
Conference held in Denver, USA, for his work on automotive radar target
detection using neural networks.
\end{IEEEbiography}

\begin{IEEEbiography}[{\includegraphics[width=1in,height=1.25in,clip,keepaspectratio]{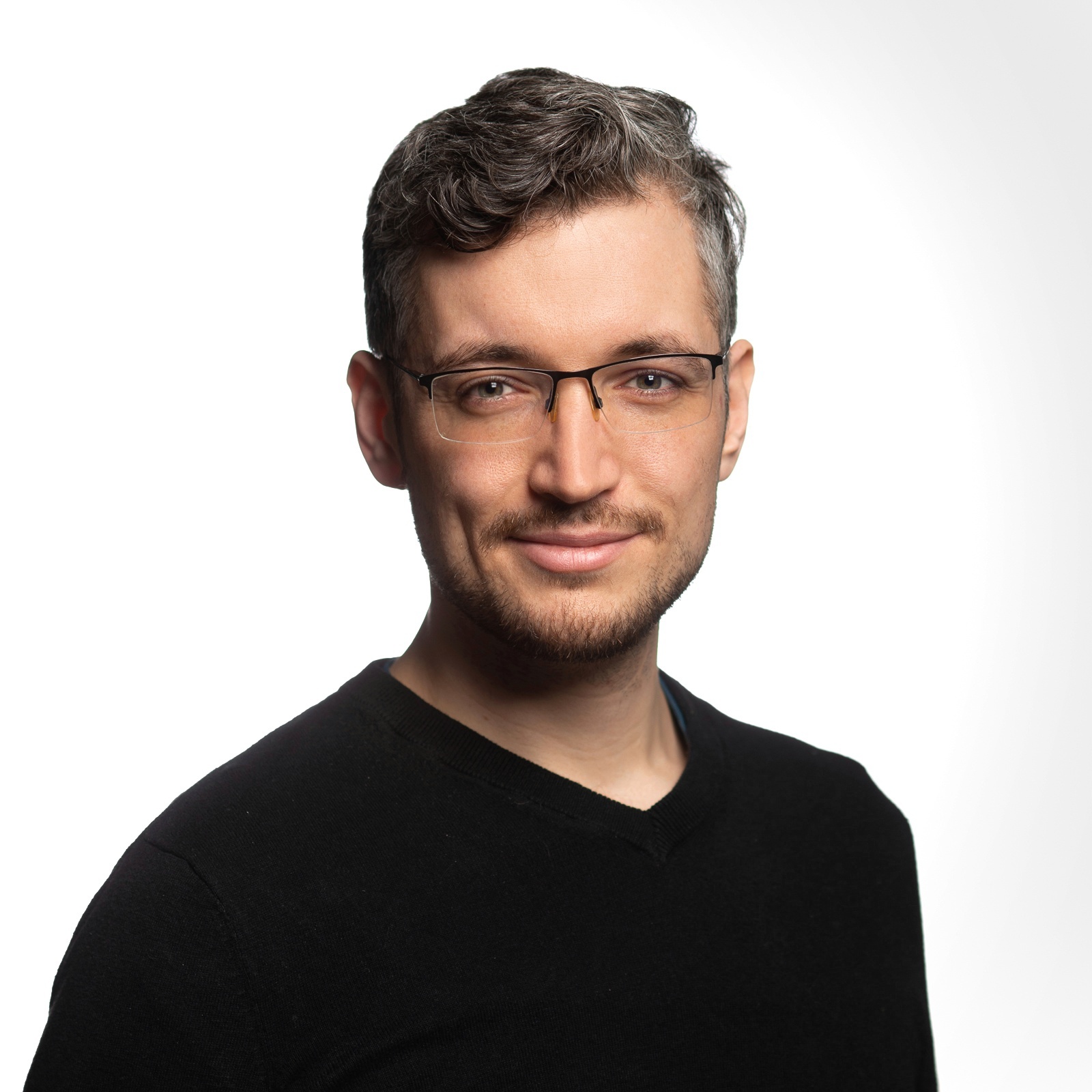}}]{Andras Palffy}
(Member, IEEE) received the M.Sc. degree in computer science engineering from Pazmany Peter Catholic University, Budapest, Hungary, in 2016, and the M.Sc. degree in digital signal and image processing from Cranfield University, Cranfield, U.K., in 2015. From 2013 to 2017, he was an algorithm researcher at Eutecus, a US based startup developing computer vision algorithms for traffic monitoring and driver assistance applications. He obtained his Ph.D. degree in 2022 at Delft University of Technology, Delft, Netherlands, focusing on radar based vulnerable road user detection for automated driving. In 2022 he co-founded Perciv AI, a machine perception startup developing AI-driven, next generation machine perception for radars.
\end{IEEEbiography}

\begin{IEEEbiography}[{\includegraphics[width=1in,height=1.25in,clip,keepaspectratio]{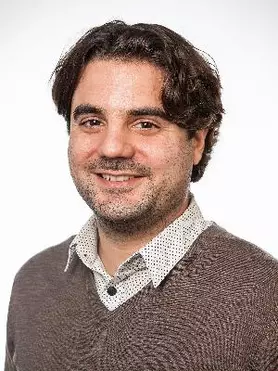}}]{Julian F. P. Kooij}
(Member, IEEE) received the Ph.D. degree in artificial intelligence from the University of Amsterdam, Amsterdam, Netherlands, in 2015. In 2013, he was with Daimler AG worked on path prediction for vulnerable road users. In 2014, he joined Computer Vision Lab, Delft University of Technology (TU Delft), Delft, Netherlands. Since 2016, he has been with Intelligent Vehicles Group, part of the Cognitive Robotics Department, TU Delft, where he is currently an Associate Professor. His research interests include probabilistic models and machine learning techniques to infer and anticipate critical traffic situations from multi-modal sensor data.
\end{IEEEbiography}

\begin{IEEEbiography}[{\includegraphics[width=1in,height=1.25in,clip,keepaspectratio]{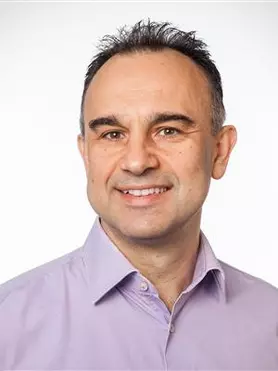}}]{Dariu M. Gavrila}
(Member, IEEE) received the Ph.D. degree in computer science from the University of Maryland, College Park,MD, USA, in 1996. From 1997, he was with Daimler R\&D, Ulm, Germany, where he became a Distinguished Scientist. 2016, he moved to Delft University of Technology, Delft, Netherlands, where he since Heads the Intelligent Vehicles Group as a Full Professor. His research interests include sensor-based detection of humans and analysis of behavior, recently in the context of the self-driving cars in urban traffic. He was the recipient of the Outstanding Application Award 2014 and the Outstanding Researcher Award 2019, from the IEEE Intelligent Transportation Systems Society.
\end{IEEEbiography}

\begin{IEEEbiography}[{\includegraphics[width=1in,height=1.25in,clip,keepaspectratio]{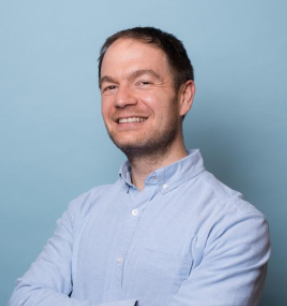}}]{Francesco Fioranelli}
(M'15–SM'19) received the Ph.D. degree with Durham University, Durham, UK, in 2014. He is currently an Associate Professor at TU Delft, The Netherlands, and was an Assistant Professor with the University of Glasgow (2016–2019), and a Research Associate at University College London (2014–2016). 

His research interests include the development of radar systems and automatic classification for human signatures analysis in healthcare and security, drones and UAVs detection and classification, and automotive radar. He has authored over 190 peer-reviewed publications, edited the books on “Micro-Doppler Radar and Its Applications” and "Radar Countermeasures for Unmanned Aerial Vehicles" published by IET-Scitech in 2020, received four best paper awards and the IEEE AESS Fred Nathanson Memorial Radar Award 2024.
\end{IEEEbiography}

\begin{IEEEbiography}[{\includegraphics[width=1in,height=1.25in,clip,keepaspectratio]{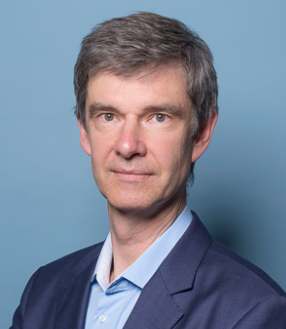}}]{Alexander Yarovoy}
(FIEEE'15) graduated from the Kharkov State University, Ukraine, in 1984 with the Diploma with honor in radiophysics and electronics. He received the Candidate Phys. \& Math. Sci. and Doctor Phys. \& Math. Sci. degrees in radiophysics from the same university in 1987 and 1994, respectively. 
In 1987 he joined the Department of Radiophysics at the Kharkov State University as a Researcher and became a Full Professor there in 1997. From September 1994 through 1996 he was with Technical University of Ilmenau, Germany as a Visiting Researcher. Since 1999 he is with the Delft University of Technology, the Netherlands. Since 2009 he leads there a chair of Microwave Sensing, Systems and Signals. 
His main research interests are in high-resolution radar, microwave imaging and applied electromagnetics (in particular, UWB antennas). He has authored and co-authored more than 600 scientific or technical papers, eleven patents and fourteen book chapters. He is the recipient of the European Microwave Week Radar Award for the paper that best advances the state-of-the-art in radar technology in 2001 (together with L.P. Ligthart and P. van Genderen) and in 2012 (together with T. Savelyev). In 2023 together with Dr. I.Ullmann, N. Kruse, R. Gündel and Dr. F. Fioranelli he got the best paper award at IEEE Sensor Conference. In 2010 together with D. Caratelli Prof. Yarovoy got the best paper award of the Applied Computational Electromagnetic Society (ACES). 
In the period 2008-2017 Prof. Yarovoy served as Director of the European Microwave Association (EuMA). He is and has been serving on various editorial boards such as that of the IEEE Transaction on Radar Systems. From 2011 till 2018 he served as an Associated Editor of the International Journal of Microwave and Wireless Technologies. He has been member of numerous conference steering and technical program committees. He served as the General TPC chair of the 2020 European Microwave Week (EuMW’20), as the Chair and TPC chair of the 5th European Radar Conference (EuRAD’08), as well as the Secretary of the 1st European Radar Conference (EuRAD’04). He served also as the co-chair and TPC chair of the Xth International Conference on GPR (GPR2004). 
\end{IEEEbiography}

\vfill

\end{document}